\documentclass[aps,prl,twocolumn,groupedaddress,showpacs,amsmath]{revtex4-1}

\usepackage{graphicx}

\begin{document}


\title{Resonant Few-Photon Excitation of a Single-Ion Oscillator}

\author{Y.-W.~Lin}
\author{S.~Williams}
\author{B.~C.~Odom}
\email[]{b-odom@northwestern.edu}
\affiliation{Department of Physics and Astronomy, Northwestern University, Evanston, IL 60208, USA}

\date{\today}

\begin{abstract}
We study the motion of an undamped single-ion harmonic oscillator, resonantly driven with a pulsed radiation pressure force. We demonstrate that a barium ion, initially cooled to the Doppler limit, quickly phase locks to the drive and builds up coherent oscillations above the thermal distribution after scattering of order one hundred photons. In our experiment, this seeded motion is subsequently amplified and then analyzed by Doppler velocimetry. Since the coherent oscillation is conditional upon the internal quantum state of the ion, this motional excitation technique could be useful in atomic or molecular single-ion spectroscopy experiments, providing a simple protocol for state readout of non-fluorescing ions with partially closed-cycle transitions.
\end{abstract}



\maketitle

The high quality of environmental isolation, storage time, and particle localization provided by ion traps creates an excellent environment for quantum control and precision spectroscopy \cite{leibfried2003}.  To date, nearly all few-ion experiments have accomplished control and addressing by relying on laser-accessible closed-cycle optical transitions, which occur in only a small number of atomic ion species.  An essential feature required for extending the power of ion traps to other atomic and molecular species is the ability to perform non-desctructive internal state readout with a small or vanishing numbers of scattered photons \cite{schmidt2006, leibfried2012, ding2012, mur2012, clark2010}.  Ability to control and monitor new atomic and molecular ion species could open new possibilities in such areas as quantum information processing, parity-violation studies, search for time-reversal symmetry breaking, and search for time-variation of fundamental constants.

Here, we study the excitation of a Doppler-cooled single barium (Ba$^+$) ion in a harmonic trapping potential with negligible damping, under the influence of a resonantly pulsed radiation pressure force in the regime of few photon scattering.  A pulsed radiation pressure force, in the regime of large scattering numbers, has previously been used for few-ion mass spectrometry \cite{drewsen2004}.  In modeling and experiment, we find that phase-locking behavior allows efficient energy transfer to the ion oscillator, such that scattering of order one hundred photons effectively separates the driven velocity from the Doppler-cooled distribution.  Since motional excitation by photon scattering is conditional upon the ion's internal state, a pulsed radiation pressure force could be used to transfer internal state information from a molecular or atomic spectroscopy ion with only a marginally closed-cycle transition onto a co-trapped logic ion used for state readout. State readout of a non-cycling spectroscopy ion has been accomplished by first reducing its temperature by laser-cooling a co-trapped logic ion, then using sophisticated protocols to map its internal state information onto detectable motion of the two-ion crystal \cite{schmidt2005, hume2011}. A pulsed radiation pressure force applied to a molecular or atomic spectroscopy ion with only a marginally closed-cycle transition could offer a new and simple means to accomplish the state mapping.

Because the driven oscillator phase locks such that photon scattering transfers maximum energy (with ion velocity along the laser direction at time of scattering,) our approach of excitation by a pulse train with small duty cycle $D$ is more efficient per photon than excitation by a sinusoidally modulated radiation pressure force. For the moment neglecting stochastic aspects of photon scattering, the driving force $\lambda$ is given by $\lambda = \hbar k \rho_e \Gamma /\sqrt{2}$, where $k$ is the photon wave number, $\rho_e$ is the excited state population, and $\Gamma$ is the scattering rate. The $\sqrt{2}$ factor accounts for the $45^{\circ}$ angle between the force and the motion in our experiment; $\lambda\sim 10^{-20}\;\text{N}$ for typical visible dipole transitions, with $\rho_e=0.3$ determined experimentally. We treat the ion as an undamped oscillator because damping from radiation and resistive losses are negligible, and damping due to off-resonant photon scattering is also unimportant, as the ion motion is small in our experiment. The dynamics can be studied by several methods such as Green's theorem. For a resonantly pulsed drive, the evolution of the oscillation amplitude $A_n$ and the phase $\phi_n$ after $n$ driving cycles are found to be
\begin{subequations}
\label{eq_eq}
\begin{eqnarray}
\label{eq_amplitude}
\Delta A &= A_{n+1}-A_{n} &= \eta \cos \phi_{n} \\
\label{eq_phase}
\Delta \phi &= \phi_{n+1}-\phi_{n} &= -\frac{\eta}{A_n} \sin \phi_n
\end{eqnarray}
\end{subequations}
where $\eta= 2 \lambda \sin(\pi D)/m \omega^2$, for ion mass $m$ and secular frequency $\omega$. To obtain Eq.~(\ref{eq_eq}) we use the weak drive approximation $\eta \ll A_n$; in our experiment $\eta = 2.7\;\text{nm}$ and $A_0\approx 34\;\text{nm}$ for a $\text{Ba}^+$ ion at the Doppler limit. The phase constant is defined such that $\phi=0$ at the middle of the pulse. By treating Eq.~(\ref{eq_eq}) as continuous in $n$ and integrating, we find closed-form expressions:
\begin{subequations}
\label{eq_solution}
\begin{eqnarray}
\label{eq_sol_amplitude}
A_n &=& \sqrt{(A_0\sin\phi_0)^2+(\eta n+A_0\cos\phi_0)^2}\\
\label{eq_sol_phase}
\phi_n &=& -\cos^{-1}(\frac{\eta n+A_0\cos\phi_0}{A_n}).
\end{eqnarray}
\end{subequations}
The peak velocity of the oscillator after $n$ cycles is then $V_n=\omega A_n$ and can be experimentally measured by Doppler velocimetry after amplification.

\begin{figure}
\includegraphics[width=3.3in]{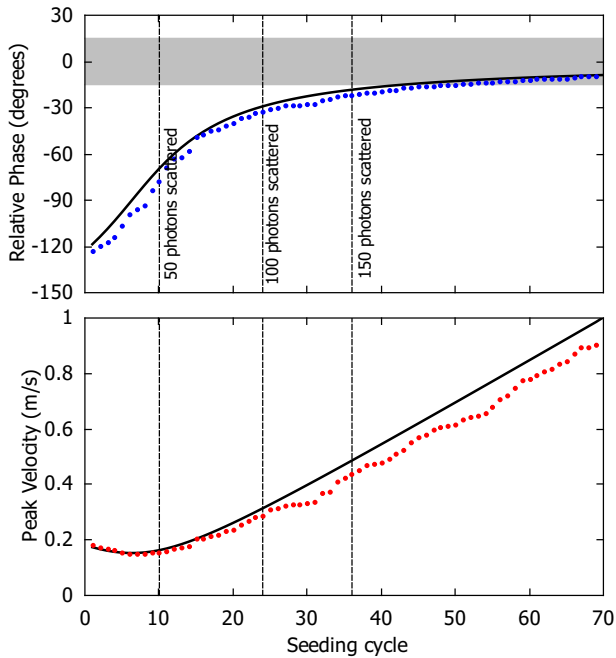}
\caption{\label{plot_sample} Modeled response of the $\text{Ba}^{+}$ oscillator to a pulsed drive with 10\% duty cycle (gray band), with realistic drive parameter $\eta=2.67\;\text{nm}$, and initial conditions $\phi_0 = -122^{\circ}$ and $A_0=34\;\text{nm}$ ($V_0=0.2\;\text{m/s}$), typical of Doppler cooling.  Results from the simulation (points) deviate from the model of Eq.~(\ref{eq_solution}) (lines) because of noise in photon scattering.}
\end{figure}

We compare Eq.~(\ref{eq_solution}) to a molecular dynamics simulation which takes the random timing and spontaneous emission angle of photon scattering into account. For the sample initial conditions and simulated scattering history shown in Fig.~\ref{plot_sample}, the simulation differs slightly from the prediction, as the randomness inputs noise into the drive strength $\eta$ and the driving phase $\phi$. We find that Eq.~(\ref{eq_sol_amplitude}) describes the ensemble average from the simulation within the experimental uncertainty. Note that scattering of order 150 photons effectively separates the $\text{Ba}^{+}$ ion velocity from the initial thermal value.

Our experimental investigation is performed with a single $\text{Ba}^{+}$ ion in a linear radio frequency trap with axial secular frequency $\omega_z = 2\pi\times926\;\text{kHz}$. We load a $^{138}\text{Ba}^+$ ion into the trap by resonance enhanced two-photon ionization with a 791.1 nm laser first driving neutral barium to the 6s6p $^3\text{P}_1$ state and a second 337 nm photon ionizing the atom \cite{steele2007}. The ion is Doppler cooled by driving the blue $\text{6S}_{1/2}\rightarrow \text{6P}_{1/2}$ transition (493.4 nm, $\Gamma_{S}=2\pi\times 15.2\;\text{MHz}$) and a red laser coupling $\text{5D}_{3/2}\rightarrow\text{6P}_{1/2}$ (649.7 nm, $\Gamma_{D}=2\pi\times 4.9\;\text{MHz}$) repumping the population. The two lasers are focused on the ion and co-propagate at $45^\circ$ with respect to the trap z-axis. For Doppler cooling, we set the blue laser intensity to $2\,I_{sat}$ with -15 MHz detuning; the red repumping laser intensity is $10\,I_{sat}$, detuned by -20 MHz.

In order to detect few-photon seeded motion of the trapped ion, we use a motional amplification scheme, where a continuous-wave (CW) laser is blue-detuned from a cycling transition \cite{vahala2009, kaplan2009}.  This technique has been used to detect motion induced by electrostatic \cite{knunz2010}, radiation pressure \cite{sheridan2012}, and optical dipole \cite{hume2011} forces, multiplying seeded velocities to amplitudes large enough for detection by Doppler velocimetry \cite{berkeland1998, biercuk2010}. Fluorescence from the ion is detected by a photomultiplier tube (PMT) with an overall efficiency of 0.1\% including a bandpass filter transmitting only the blue light. We use an FPGA-based counter to perform photon statistics.

\begin{figure}
\includegraphics[width=3.3in]{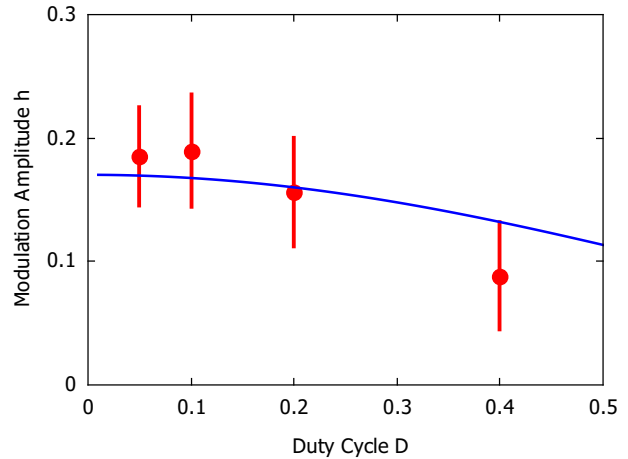}
\caption{\label{plot_dcycle} Measured modulation amplitude mean (points) and standard deviation (bars) versus seeding duty cycle, with seeding time varied to maintain $D\times t_s=4\;\mu\text{s}$; data were collected over 30 trials. Averaging Eq.~(\ref{eq_solution}) over initial phases and fitting for amplification-stage gain yields the solid curve.}
\end{figure}

The major cycle of the experiment consists of seeding the motion, amplification, and resetting by Doppler cooling. To seed the ion motion, we chop the blue laser at the secular frequency $\omega_z$; the laser intensity is set to $10\,I_{sat}$ with zero detuning; the seeding pulse train is applied for time $t_s$. The blue laser detuning is then set to +15 MHz for time $t_a$ to amplify the seeded oscillation; $t_a=10\;\text{ms}$ was chosen by experimental optimization. After amplification, the laser detuning is then set back to -15 MHz to damp the excited ion. The repumping laser is not altered for each stage of the experiment. To detect the motion, we collect fluorescence from the last 4 ms of the amplification stage and the first 4 ms of the cooling stage. The above experiment cycle is repeated every 50 ms, and we integrate for 2 seconds to obtain the modulated fluorescence signal. These 40 cycles typically yield 1000 photon counts, collected by the FPGA into 20 timing bins of width 46.7 ns. Ion motion is detected as a modulation in photon arrival times with modulation amplitude $h$ determined by fitting the correlation function $g(\tau )$ to $1+h\cos(\omega_z (\tau-\tau_0))$ where $\tau$ is the time referenced to each secular motion cycle, and $\tau_0$ compensates for constant experimental phase delays.

\begin{figure}
\includegraphics[width=3.3in]{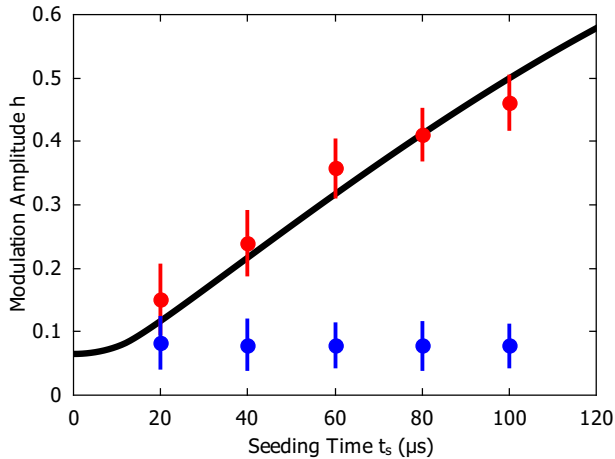}
\caption{\label{plot_time} Modulation amplitude versus seeding time, when the seeding pulses excite an S-state ion (red), and when the ion is shelved in the D-state (blue). Each point is the average of 30 measurements, with the vertical bars showing the distribution standard deviation (rather than the error on the mean). The predicted response (black curve) is from  Eq.~(\ref{eq_solution}), fitting for amplification-stage gain.}
\end{figure}

We first investigate the effect of pulse width on seeding, maintaining a constant average number of scattered photons. Fig.~\ref{plot_dcycle} shows the modulation amplitude for different seeding pulse duty cycles, with the seeding time $t_s$ is adjusted to keep fixed the laser-ion interaction time $D\times t_s$. We observe a stronger excitation for a shorter duty cycle as predicted by Eq.~(\ref{eq_solution}) normalized to scattering number. The fitting model is an ensemble average over the initial phases $\phi_0$. We treat the coherent amplification as a constant velocity gain $g_a$ (here $g_a = 2$), found from a single-parameter fit to the data, and then convert the amplified ion velocity into modulation amplitude according to a Lorentzian spectrum using an experimentally determined full-width-half-maximum of 30 MHz. Deviation between the experiment and the theory is attributed to known variation of the amplification gain as the laser frequencies drift during the experiment.

Fig.~\ref{plot_time} shows the measured modulation amplitude versus seeding time, using $D=0.1$, with the predicted response from Eq.~(\ref{eq_solution}) and a single-parameter fit for amplification gain (here $g_a = 2.7$.) Due to the initial distribution of oscillator phases, the model predicts slow ensemble-averaged velocity buildup at short times, until some degree of phase locking has occurred (see also Fig.~\ref{plot_sample}.) The time scale for phase-locking is given by $A_0/\eta\approx 13$ cycles, or $14\;\mu\text{s}$ in this work. Once the phase is locked, all the photon momentum contributes to secular motion excitation coherently. Since the oscillation converts to fluorescence modulation approximately linearly, modulation amplitude goes linearly with the seeding time for a phase-locked oscillator.

\begin{figure}
\includegraphics[width=3.3in]{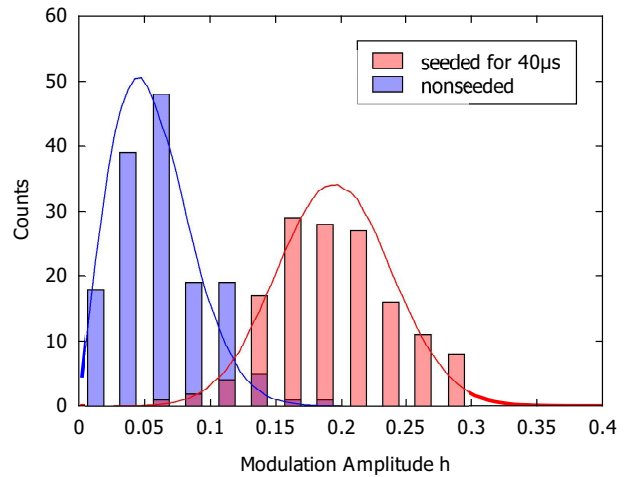}
\caption{\label{plot_histogram} Distribution of the modulation amplitudes, measured after seeding the motion for $40\;\mu\text{s}$ (red histogram) and for an unseeded ion (blue histogram). The simulation (solid curves) accounts for noise in ion dynamics and shot noise in detection. Amplification gain is the single fit parameter.}
\end{figure}

We now consider the application of our pulsed seeding technique to internal state readout by mapping the internal state onto ion motion. Fig.~\ref{plot_time} also shows the response from an unseeded ion, obtained by optically pumping into the $\text{D}_{3/2}$ state before seeding pulses are applied; before amplification the ion is repumped to the $\text{S}_{1/2}$ state. In the context of a spectroscopy experiment, (failed) seeding from the $\text{D}_{3/2}$ state simulates state readout after a successfully driven spectroscopy transition, while (successful) seeding from the $\text{S}_{1/2}$ state simulates state readout after a failed spectroscopy transition (or vice versa.) Based on the magnitude of the dark state baseline, seeding for $t_s\approx 40\;\mu\text{s}$ is sufficient to create Ba$^+$ motion well separated from the noise floor after amplification and our integration over 40 excitation/detection cycles. For this seeding time, there are approximately $n_\gamma = 150$ photons scattered per excitation cycle, and the measured fluorescence modulation is $\bar{h}=0.24$. In Fig.~\ref{plot_histogram} we show the experimental $h$ distributions for seeded and unseeded ions, again using $D=0.1$, along with the simulated results (with $g_a = 2.5$ the only fit parameter) accounting for various noise sources summarized in Table \ref{table_error}.

\begin{table}
\caption{\label{table_error} Modeled contribution of various noise sources to the distribution width of the modulation amplitude, for $n_\gamma = 150$ yielding $\bar{h}=0.20$. Data is integrated over $N$ excitation/detection cycles, with initial ion temperature $360\;\mu\text{K}$, $t_s=40\;\mu\text{s}$, $t_a=10\;\text{ms}$, and $g_a=2.5$. For $N=40$, there are 1000 fluorescence photon counts spread over 20 timing bins. The last line represents the quadrature addition of all sources.}
\begin{ruledtabular}
\begin{tabular}{lcc}
Source & $\Delta h$, $N=1$\footnote{Considering only the noise intrinsic to ideal single-shot excitation, e.g. for perfect fluorescence collection or for sustained oscillation during detection.} & $\Delta h$, $N=40$ \\
\hline
Photon counting & 0 & 0.043 \\
Initial thermal motion & 0.072 & 0.011 \\
Seeding stage & 0.032 & 0.005 \\
Amplification stage& 0.114 & 0.018\\
\hline
Total width & 0.139 & 0.048 \\
\end{tabular}
\end{ruledtabular}
\end{table}

In our experimental implementation, the major source of noise is low photon detection efficiency. Photon shot noise propagates through the data analysis and contributes to the width. In addition to the seeding noise already discussed, the coherent amplification processes injects noise into the ion motion, as the photon scattering has random spatial and temporal components. In Table \ref{table_error} we model this noise term as a random walk in velocity space, $(\Delta v)^2\approx v_r^2 \rho_e \Gamma t_a$, where $v_r=\hbar k/m$ is the recoil velocity and $\rho_e\approx 0.02$ during amplification. Overall, the summarized error sources characterize our state-detection uncertainty and form the distribution in Fig.~\ref{plot_histogram}. If we discriminate whether the ion's motion is excited by a threshold value $h_{th}=0.12$, where the two distributions intersect, the false positive rate is estimated to be 3.2\% and the false negative rate is 3.8\%.

Note that the measurement uncertainty is reduced by $N^{-1/2}$ after integrating over $N$ excitation/detection cycles. ($N=40$ in this work.) However, the required number of photons for state discrimination is then $N\times n_\gamma$, where $n_\gamma\approx 150$ is the seeding photon number required here. Currently, we perform detection while the ion velocity is either being amplified or cooled.  An important improvement, which would eliminate the need to integrate over $N > 1$ excitation cycles, would be to simultaneously damp and amplify the oscillation with both the cooling and the repumping laser to achieve self-limiting sustained large-amplitude oscillation \cite{vahala2009, knunz2010}. The number of scattered photons required for seeding the motion could also be reduced by a factor of $\sqrt{2}$ by sending the laser along the trap axis, making the photon momentum parallel to the secular motion direction. The detection sensitivity could also be enhanced by fixing the modulation phase (equivalent to a lock-in technique) rather than fitting for it, and by increasing the photon detection efficiency \cite{streed2011, shu2010}. Seeding a lighter ion would require fewer scattered photons, as the ratio of recoil to thermal velocity goes as $m^{-1/2}$.

If the trap is loaded with both a spectroscopy ion (used for the seeding stage) and a fluorescing logic ion (used first to Doppler cool the two-ion system then later in the amplification/detection stage), our state discrimination method can be applied to spectroscopy experiments. For instance, in our lab we are pursuing silicon monoxide ion ($\text{SiO}^{+}$) spectroscopy \cite{nguyen2011pra}, where a single $\text{SiO}^{+}$ ion is co-trapped and sympathetically cooled by a $\text{Ba}^{+}$ ion. For $\text{SiO}^{+}$ the seeding laser would drive the dipole transition between $|X^{2}\Sigma^{+},\; v=0\rangle$ and $|B^{2}\Sigma^{+},\;v=0\rangle$ ($\lambda =385\;\text{nm}$) \cite{cameron1995, cai1999}. Before the population decays into the $v=1$ vibrational manifold, there are on average 150 $B$-$X$ scatters. The seeding fidelity can be improved by repumping from $v=1$, allowing on average over 7000 scatters before decay to $v=2$ or to the low-lying $A^{2}\Pi$ manifold. Further study of seeding behavior in a 2-ion crystal is required, but we expect approximately a factor of 2 degradation from the 1-ion seeding efficiency, to account for excitation of additional non-detected normal modes.  With a vibrational repump, our simple seeding technique should be suitable for $\text{SiO}^{+}$ spectroscopy readout, and with improvements to the amplifier stage discussed above, the vibrational repump might not be required.

It is instructive to compare this state readout approach to other protocols using a co-trapped spectroscopy and logic ion. Quantum logic spectroscopy  \cite{schmidt2005} does not use spectroscopy ion scattering, but places restrictions on transition linewidth and wavelength, also requiring ground state cooling and logic ion shelving. Coherent excitation by an optical dipole force \cite{hume2011} also does not scatter from the spectroscopy ion and relaxes ground-state cooling and shelving constraints, but it requires a suitable transition and challenging alignment of counter-propagating beams onto a dark ion.  Our pulsed-excitation method does not require ground state cooling or shelving, can in principle be applied to any transition, and is quite simple to implement; however, it is limited to spectroscopy species with partially closed-cycle transitions allowing repeated scattering. Finally, sympathetic heating spectroscopy \cite{clark2010} uses non-modulated spectroscopy ion scattering, requiring many more scattered photons than the phase-coherent approach described here.

To conclude, we have modeled and experimentally studied the state-dependent excitation of a single-ion oscillator impulsively driven at the trap frequency.  Rapid phase-locking behavior results in efficient energy transfer; scattering approximately $n_\gamma = 150$ photons effectively separates the bright and dark state velocity distributions. However, our detection method is currently inefficient, requiring $N=40$ excitation/detection cycles to build up photon statistics, such that 6000 scattered photons are currently needed to determine the internal state. Implementation of sustained amplification with phase-sensitive detection, along with other technical improvements discussed above, could reduce scattering to the small number (order 100 and $N=1$) required to make a seeded excitation detectable above the thermal and scattering noise.  With a co-trapped fluorescing ion used for the amplification and detection, this motional seeding technique could be useful in spectroscopy experiments on molecular ions with semi-closed transitions \cite{lien2011, nguyen2011pra, nguyen2011njp}, atomic ions with slow cycling transitions, and in bichromatic force schemes where cycling is enhanced by stimulated emission \cite{voitsekhovich1989, grimm1990, chieda2011}. This resonant impulsive excitation technique could also be used to coherently excite selected normal modes of larger trapped ion crystals.

We wish to thank John Bollinger, Jens Koch, Jason Nguyen, and  Kerry Vahala for helpful discussions. This work is supported by AFOSR YIP (Grant No. FA9550-10-1-0221), NSF CAREER (Grant No. PHY08-47748), and the Alfred P. Sloan Foundation (Grant No.  BR-5104.)




%

\end{document}